\documentclass{hep99}
\usepackage{epsfig}

\newcommand{\beq}{\begin{equation}}
\newcommand{\eeq}{\end{equation}}
\newcommand{\bit}{\begin{itemize}}
\newcommand{\eit}{\end{itemize}}
\newcommand{\bce}{\begin{center}}
\newcommand{\ece}{\end{center}}

\newcommand{\jp}{J/\psi}
\newcommand{\cts}{\cos  \theta^{\ast}}
\newcommand{\ctts}{\cos^{2}  \theta^{\ast}}

\newcommand{\g}{\mathrm{GeV}/c}
\newcommand{\uo}{\Upsilon(1S)}
\newcommand{\up}{\Upsilon}
\begin{document}


\title{Recent results on $\jp$, $\psi(2S)$ and $\Upsilon$ production at CDF}

\author{Robert Cropp$^\dag$}

\address{Department of Physics, University of Toronto, Toronto, Canada,
 M5S 1A7\\[3pt]
$^\dag$For the CDF Collaboration\\[3pt]
E-mail: {\tt rjcropp@physics.utoronto.ca}}

\abstract{CDF has obtained new results on quarkonium production in $p\bar{p}$
collisions at $\sqrt{s} = 1.8$ TeV. We report on measurements of 
 $\Upsilon$ meson production,
  $\Upsilon(1S)$ production from $\chi_b$ 
feeddown, and the production polarization of $\Upsilon(1S)$, $\jp$ and $\psi(2S)$
mesons.} 

\maketitle

\section{Introduction}

Heavy quarkonium production provides an opportunity to study QCD, both in 
its perturbative and nonperturbative regimes.
Several analyses of quarkonia production have previously been performed
by CDF using a subset of the Run I dataset. Differential cross sections
for the production of $\jp$, $\psi(2S)$ and $\up$ mesons were measured
\cite{psi1a,ups1a}. The $\psi$ charmonia produced in $B$-hadron
decays were separated from the prompt (zero-lifetime) charmonia using 
vertex displacement. In addition, the contribution of
$\chi_c$ feeddown to prompt $\jp$ production was measured \cite{chic}.
It was found that the cross sections for direct
$\jp$ and $\psi(2S)$ production  (i.e. not from feeddown) 
were significantly higher than predicted
by the Color Singlet Model \cite{csm}, by a factor of approximately 50.
This result stimulated the inclusion
of color-octet $c\bar{c}$ states in theoretical calculations of
quarkonium production \cite{com}.

In this paper we present more recent studies of quarkonia production at CDF.
These include new $\up$ cross section measurements, the observation of
$\chi_b \rightarrow \uo \gamma$ feeddown, and
measurements of $\jp$, $\psi(2S)$ and $\uo$ production polarization.

The dataset used for these analyses consists of approximately
110 pb$^{-1}$ of $p\bar{p}$ collisions recorded during Run I
of the Tevatron in 1992-95. $\jp$, $\psi(2S)$ and $\up$ candidates
are all recorded in their decay mode to $\mu^+\mu^-$, using a 
three-level dimuon trigger. Muons in the  central pseudorapidity region
  ($|\eta|<\sim0.6$) are used. Muon candidates consist of tracks in
the central tracking chamber matched to hits in muon chambers,
located outside the calorimeter. Precise vertexing is performed by the
silicon vertex detector (SVX), which measures track impact parameters
with an asymptotic resolution of $13 \ \mu$m.
Photon candidates consist of energy deposits in the central electromagnetic calorimeter,
matched to clusters in strip chambers embedded in the calorimeter.

\begin{figure}
\begin{center}
\begin{minipage}{3.5cm}
\epsfig{file=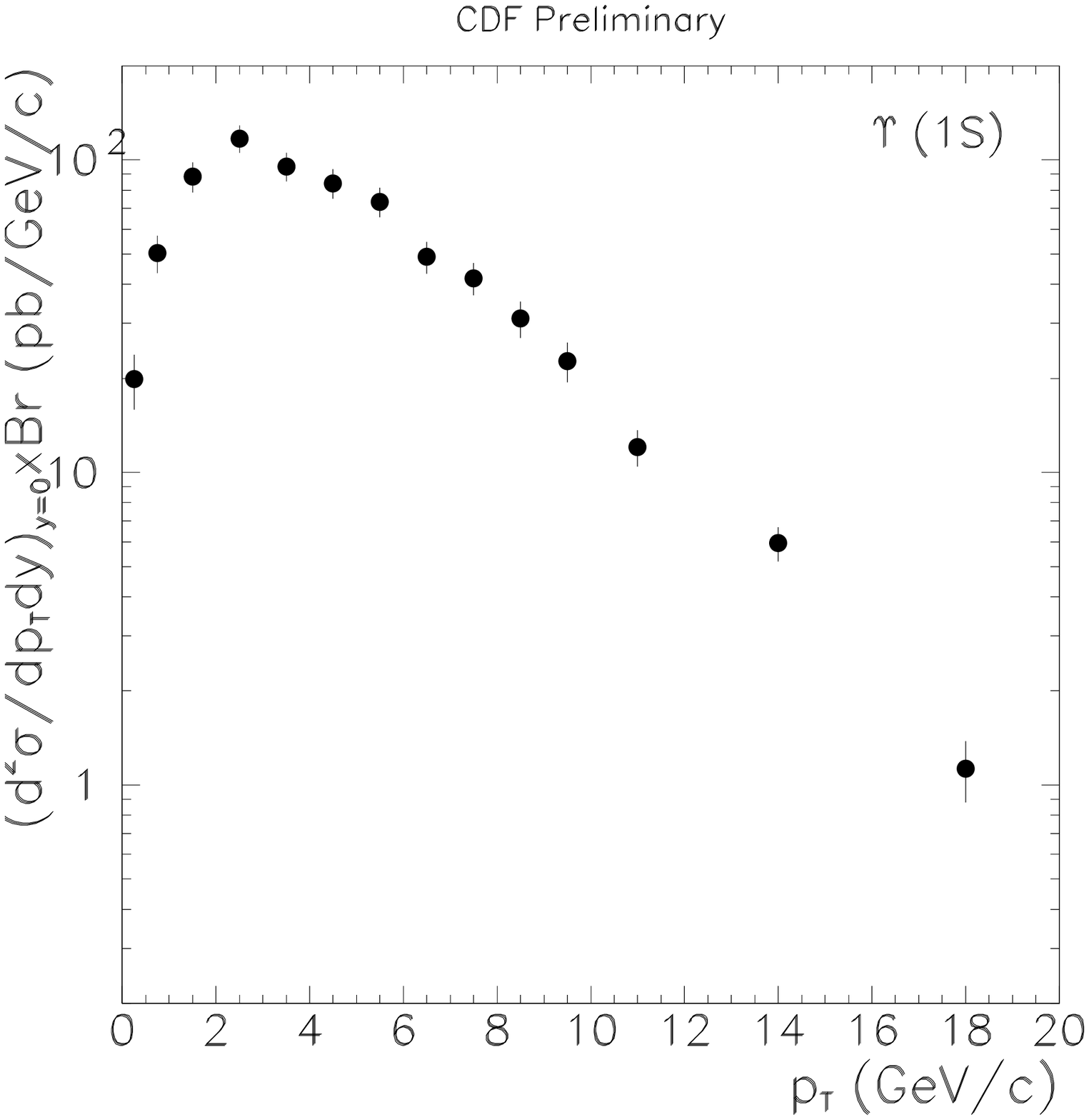,bbllx=30pt,bblly=130pt,bburx=540pt,bbury=660pt,width=3.5cm}
\end{minipage}
\begin{minipage}{3.5cm}
\epsfig{file=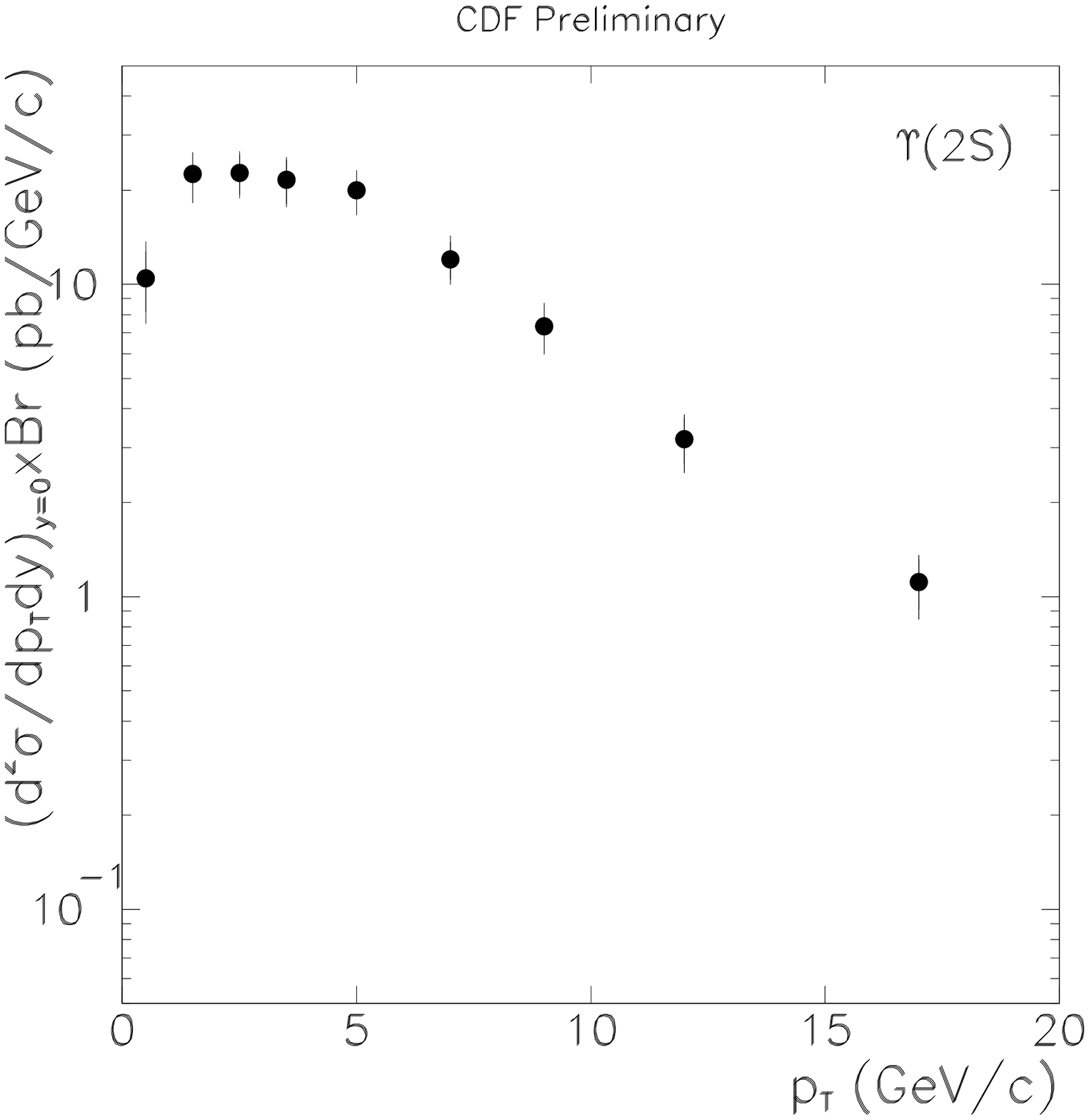,bbllx=30pt,bblly=130pt,bburx=540pt,bbury=660pt,width=3.5cm}
\end{minipage}
\end{center} 
\begin{center}
\begin{minipage}{3.5cm}
\epsfig{file=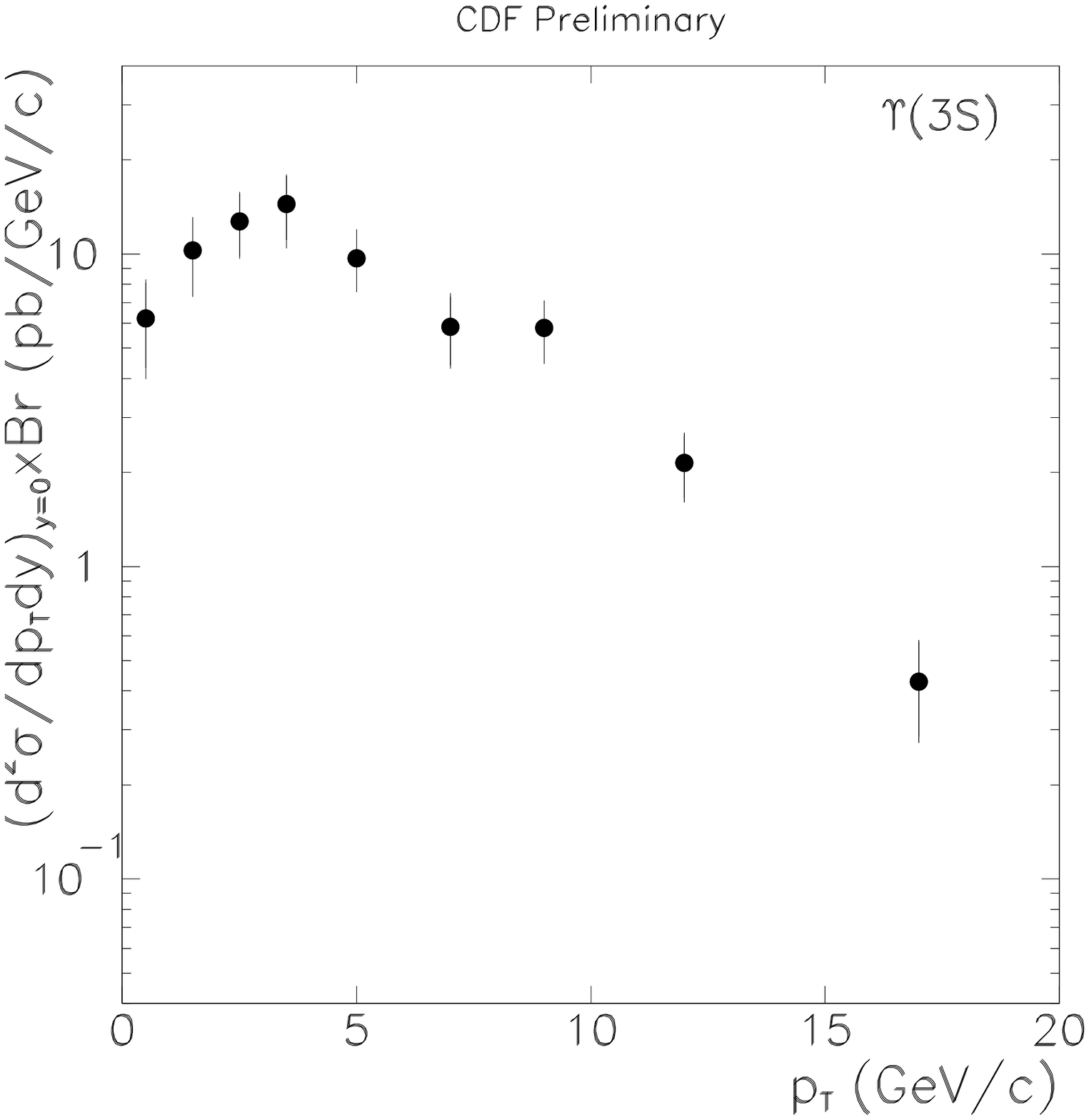,bbllx=30pt,bblly=130pt,bburx=540pt,bbury=660pt,width=3.5cm}
\end{minipage}
\begin{minipage}{3.5cm}
\epsfig{file=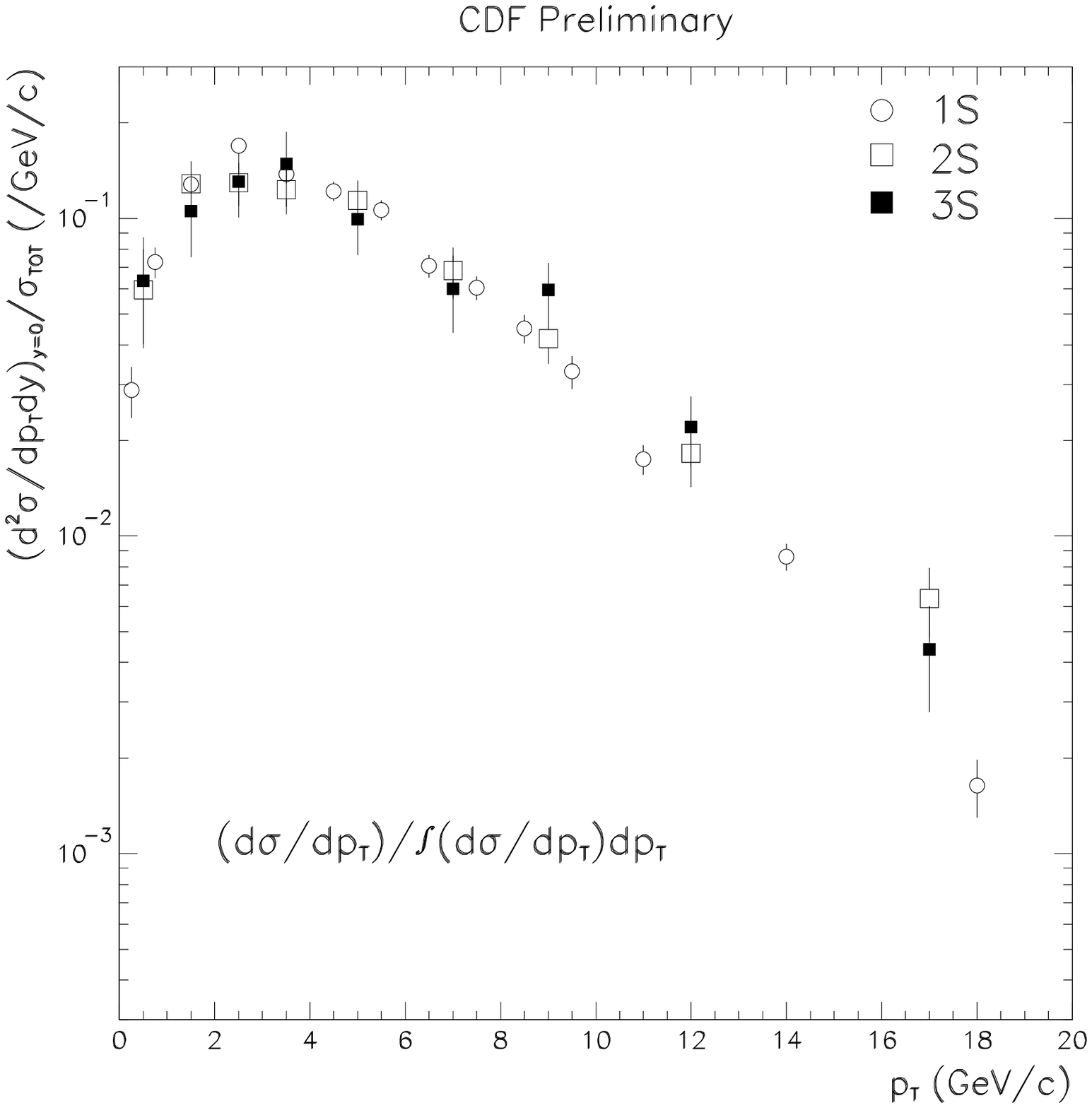,bbllx=30pt,bblly=130pt,bburx=540pt,bbury=660pt,width=3.5cm}
\end{minipage}
\end{center} 
\vspace{-5mm}
\caption{Production cross section times dimuon branching ratio for $\uo$, $\up(2S)$
and $\up(3S)$, and a comparison of their shapes.}
\label{Fxsec}
\end{figure}

\section{Production $d\sigma/dP_T$ for $\uo,\up(2S),\up(3S)$} \label{Sxsec}

The inclusive production cross section for $\up$ mesons has been measured over the
kinematic range $|y^\up| < 0.4$ and $0<P_T^\up<20 \ \g$. We use a 77 pb$^{-1}$
subset of the data, a significantly larger sample than in the previous
published result.
 $\uo$, $\up(2S)$ and $\up(3S)$ candidates are counted by fitting the dimuon
invariant mass distribution to Gaussian signals on a quadratic background. 
The acceptance and efficiency in each $P_T$ bin 
are calculated using a simulation of the
detector and trigger. The resulting cross sections are shown in Fig.~\ref{Fxsec}.
A comparison of the normalized $\uo$, $\up(2S)$ and $\up(3S)$
cross sections shows that they have similar shapes.

\section{$\uo$ production from $\chi_b$ feeddown}

\begin{figure}
\bce
\epsfig{file=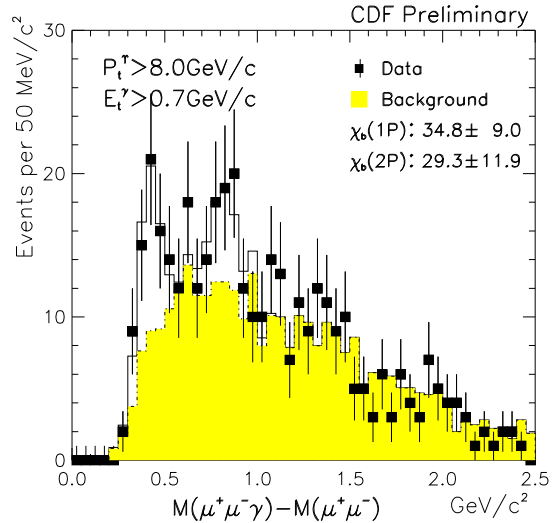,width=7.8cm}
\vspace{-7mm}
\caption{$\uo\gamma$ invariant mass distribution, showing
 $\chi_b(1P)$ and $\chi_b(2P)$ signals above background.}
\label{Fchi}
\ece
\end{figure}

The production of $\chi_b(1P)$ and $\chi_b(2P)$ mesons has been observed
by fully reconstructing their decays to $\uo + \gamma$. $\uo$ candidates
with $P_T > 8 \ \g$ are combined with photons with $E_T^\gamma > 0.7$
 GeV\footnote{Transverse energy $E_T \equiv E \cdot \sin\theta$, analogous to $P_T$.}.
The $\uo$ $P_T$ requirement removes events with large backgrounds from
 low energy photons.  
 The combined invariant mass distribution is shown in
 Fig.~\ref{Fchi}. Background to the $\chi_b$ signal is primarily due to real 
photons from $\pi^0$ and $\eta$ decays. To model the shape of the 
background, a Monte Carlo method is used: real charged particles 
in each event are ``replaced'' by $\pi^0$, $\eta$ and $K_S^0 \rightarrow
 \pi^0 \pi^0$. The photons obtained from these fake decays are passed
through a detector simulation and combined with the real $\uo$ candidates.
Clear $\chi_b(1P)$ and $\chi_b(2P)$ signals can be seen in Fig.~\ref{Fchi},
while no $\chi_b(3P)$ signal is apparent. Taking into account the acceptance
and efficiency for reconstructing the photons, we measure the fractions
of $\uo$ with $P_T > 8 \ \g$ produced by $\chi_b(1P)$ and $\chi_b(2P)$ decays
to be $(26.7 \pm 6.9 \pm 4.3)\%$ and $(10.8 \pm 4.4 \pm 1.3)\%$ respectively.
Then, subtracting the calculated fraction of $\uo$ produced by $\up(2S,3S)$ decays, 
we obtain the direct fraction of $\uo$ with $P_T > 8 \ \g$: $(51.8 \pm 8.2
 \pm^{9.0}_{6.7})\%$.

\section{$\uo$ polarization}

The production polarization of $\up$ (and $\psi$) mesons
is measured using the distribution of the decay angle $\theta^\ast$.
This is the angle between the $\mu^+$ direction in the meson rest frame and
the meson direction in the lab frame. The distribution has the form
 $1 + \alpha  \ctts$, where the polarization parameter $\alpha$ is
+1 in the case of tranverse polarization, -1 for longitudinal and 0 for unpolarized
 production.

The $\uo$ polarization has been measured over the range $|y^{\uo}| < 0.4$, by
fitting the observed distribution of $\cts$ to a sum of transverse
and longitudinal Monte Carlo templates. These templates are drawn from
generated samples of $\uo$ decays with $\alpha$ = +1 (or -1), which
have been passed through detector and trigger simulations to account
for the dependence of the acceptance on $\cts$. Using a data sample of
approximately 4,400 $\uo$ decays, the fitted longitudinal
fraction for the $P_T$ range $2-20 \ \g$ is $\Gamma_L/\Gamma = 0.37\pm0.04$,
 consistent with
no polarization. In the more restricted $P_T$ range of $8-20 \ \g$, the
fitted fraction is $\Gamma_L/\Gamma = 0.32\pm0.11$. These values represent the
inclusive $\uo$ polarization, as it is not possible to separate direct
production from feeddown in this measurement.

Measuring the production polarization has also significantly
reduced the systematic uncertainty on the cross section measurement of
section \ref{Sxsec}.

\section{Prompt $\jp$ polarization}

The $\jp$ polarization has been measured over the kinematic range 
 $|y^{\jp}| < 0.6$, using approximately 180,000 $\jp$ decays.
 The polarization is measured both for promptly produced $\jp$ mesons
and for those produced in $B$-hadron decays.
The values of $\alpha_{Prompt}$ and $\alpha_B$ are measured in seven $P_T$ bins, 
covering $4-20 \ \g$. As with the $\uo$, it is not feasible to separate the
feeddown (from $\chi_c$ and $\psi(2S)$ decays), which accounts for about 35\%
of the prompt component \cite{chic}.

The precise vertex tracking done by the SVX is used to separate 
the prompt and $B$-decay components.
 The two muon tracks are fitted to a common
vertex, and the displacement of this vertex from the beamline is converted into
an estimate of $ct$, the proper lifetime of the decay. Prompt $\jp$ candidates have
$ct$ consistent with zero, whereas those from $B$ decay have an exponential $ct$
distribution. $\jp$ candidates are divided into short-lived and long-lived samples based
on their $ct$.
These two samples are dominated by prompt production and $B$-decay respectively, and
 the actual fractions of prompt and $B$-decay in each are measured using
a maximum-likelihood lifetime fit.

\begin{figure}
\bce
\epsfig{file=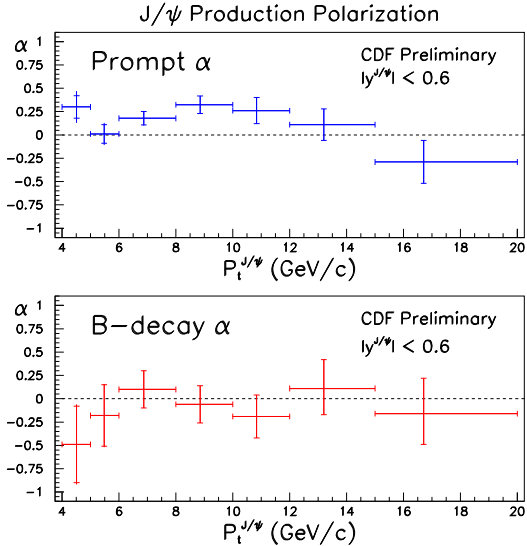,width=7.8cm}
\vspace{-9mm}
\caption{The polarization parameter $\alpha$ as a function of $P_T$, for
prompt $\jp$ mesons and for $\jp$ mesons from $B$-hadron decay.}
\label{Fjpp}
\ece
\end{figure}

The polarization fit, as in the $\uo$ case, uses Monte Carlo templates
to account for the detector and trigger acceptance. The $\cts$ distributions
in the short- and long-lived samples are simultaneously fitted for $\alpha_{Prompt}$
and $\alpha_{B}$. A third sample, in which
the muons do not have SVX information, is also included in the fit to improve the
overall precision on $\alpha$. The fit results are shown in Fig.~\ref{Fjpp}. The prompt
polarization does not display a significant increase at high $P_T$. Although
$\alpha_{Prompt}$ includes contribution from feeddown, this result does not seem
to support predictions of predominantly transverse polarization given in \cite{pol}.

\section{Prompt $\psi(2S)$ polarization}

\begin{figure}
\bce
\epsfig{file=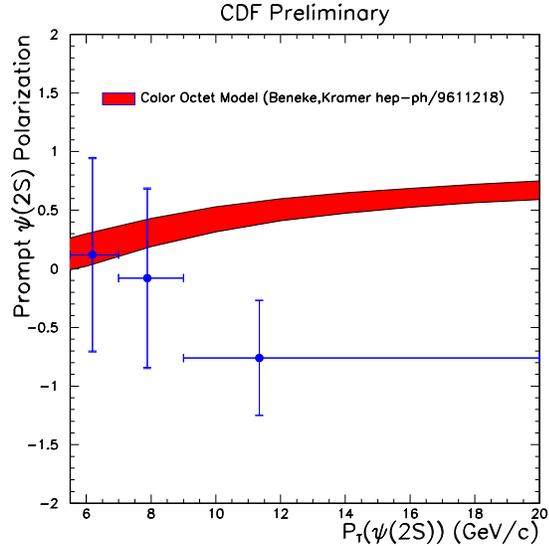,width=7.8cm}
\vspace{-9mm}
\caption{The polarization parameter $\alpha$ as a function of $P_T$ for
prompt $\psi(2S)$ mesons, compared to a color octet model prediction.}
\label{Fpsi2}
\ece
\end{figure}

The procedure for measuring prompt $\psi(2S)$ polarization is largely
similar to that in the $\jp$ case. However, this measurement is more statistically
limited, as there are about 1,800 $\psi(2S) \rightarrow \mu^+\mu^-$ candidates.
Since the $\psi(2S)$ incurs no feeddown from heavier $c\bar{c}$ states, the
prompt polarization (shown in Fig.~\ref{Fpsi2}) 
is equivalent to the direct polarization.
At high $P_T$, our measurement does not appear to support the 
model's prediction.

\section{Summary}

We have presented recent measurements of $\up$ cross sections, $\chi_b \rightarrow
\uo$ feeddown, and $\uo$, $\jp$ and $\psi(2S)$ polarization. We look forward to
the use of this information in quarkonium production models.

\end{document}